\documentclass[preprint,prd,amsmath,amssymb,nofootinbib,floatfix]{revtex4-2}

\usepackage{graphicx}
\usepackage{bm}
\usepackage{bbm}
\usepackage{physics}   
\usepackage{tensor}
\usepackage{xcolor}
\usepackage{booktabs}
\usepackage{amsthm}
\usepackage{hyperref}  
\hypersetup{colorlinks=true, linkcolor=blue, citecolor=blue, urlcolor=blue}

\newcommand{\ixi}{\iota_\xi}
\newcommand{\Lxi}{\mathcal{L}_\xi}
\newcommand{\vol}{\sqrt{-g}}
\newcommand{\Ecan}{E_{\mathrm{can}}}
\newcommand{\Smerge}{\Delta S_{\mathrm{merge}}}
\newcommand{\Stotal}{\Delta S_{\mathrm{total}}}
\newcommand{\Srad}{\Delta S_{\mathrm{rad}}}
\newcommand{\rhp}{r_{\mathrm{hp}}}
\newcommand{\Ehpm}{E^{-}_{\mathrm{hp}}}
\newcommand{\Ehpp}{E^{+}_{\mathrm{hp}}}

\newcommand{\PhiH}{\Phi_{H}}

\newcommand{\GMG}{GMGHS}
\newcommand{\EMD}{EMD}
\newcommand{\WCCC}{WCCC}
\newcommand{\WGC}{WGC}
\newcommand{\SDC}{SDC}
\newcommand{\epsH}{\tilde{\epsilon}}

\begin{document}

\title{Revisit Weak Cosmic Censorship in
       Einstein--Maxwell-Dilaton Black Holes:\\
       Third-Order Protection, Swampland Distance Conjecture,
       and Weak Gravity Conjectures}

\author{Sheng-Hong Lai}
\email{shlai@cycu.edu.tw}
\affiliation{Department of Physics and Center for High Energy Physics,
             Chung Yuan Christian University, Chung Li City, Taiwan}

\author{Wen-Yu Wen}
\email{wenw@cycu.edu.tw}
\affiliation{Department of Physics and Center for High Energy Physics,
             Chung Yuan Christian University, Chung Li City, Taiwan}

\preprint{CYCU-HEP-26-07}
\begin{abstract}
We extend the earlier analysis of cosmic censorship and the Weak Gravity
Conjecture (\WGC) in Einstein--Maxwell-dilaton (\EMD) theory~\cite{Yu:2018}
by performing a complete higher-order Sorce--Wald gedankenexperiment on the
static charged Gibbons--Maeda--Garfinkle--Horowitz--Strominger (\GMG) black
hole.
In the probe limit, an extremal \GMG\ black hole can apparently be overcharged
by a test particle, seemingly violating the weak cosmic censorship conjecture
(\WCCC)~\cite{Yu:2018}, which is nothing but first-order variational identity in the Iyer--Wald formalism~\cite{Sorce:2017}.  
This window can be closed by either incorporating backreaction via the Hoop Conjecture~\cite{Yu:2018} or further including second-order variational identities~\cite{Jiang:2019}.  In this paper, we explicitly compute third-order variational identities.  The extremality parameter $\eta(\lambda)=2M(\lambda)^{2}-Q(\lambda)^{2}$ remains non-negative to the third perturbative orders confirming \WCCC\ protection, regardless of the apparent failure of all-order generalization, model-independent proof of Lu, Wu and L\"u~\cite{Lu:2025}.
We further incorporate the Swampland Distance Conjecture (\SDC): the
resulting infinite tower of light charged states discharges any would-be
overcharged configuration on a timescale $\tau_{\mathrm{discharge}}\ll 2M$,
exponentially faster than a horizon can be destroyed, reinforcing \WCCC\ in
dynamical regimes and strongly relaxing the original \WGC\ charge-to-mass
bound.
\end{abstract}

\maketitle
\bigskip

\section{Introduction}
\label{sec:intro}

The weak cosmic censorship conjecture (\WCCC)~\cite{Penrose:1969,Penrose:1965}
asserts that singularities formed by gravitational collapse are shielded from
distant observers by event horizons, thereby preserving the predictive
character of classical general relativity.
In a landmark thought experiment, Wald~\cite{Wald:1974} demonstrated that an
extremal Kerr--Newman black hole cannot be overcharged or overspun by a test
particle, because the very particle required to violate the bound is
gravitationally repelled before it can be absorbed.
This gedankenexperiment has since been extended in many directions.
For nearly extremal Reissner--Nordstr\"om (RN) and Kerr black holes, apparent
violations at linear order~\cite{Hubeny:1999} were shown to be repaired once
second-order backreaction and self-force effects are included, as demonstrated
rigorously by Sorce and Wald~\cite{Sorce:2017} using the Iyer--Wald covariant
phase-space formalism.

In Ref.~\cite{Yu:2018}, we examined \WCCC\ in the
Einstein--Maxwell-dilaton (\EMD) theory with dilaton coupling $a=1$, whose
black hole solutions are the \GMG\ family~\cite{Garfinkle:1991}.
In the probe limit, an extremal charged dilaton black hole can apparently be
destroyed by a test particle carrying energy within specific window.
Nevertheless, the Hoop Conjecture is treated as a proxy for backreaction and closes
the allowed energy window.  On the other hand, the Generalized Second Law, enforced
during the two-step merging process across the hoop radius, imposes a charge-to-mass lower bound
$q/m\gtrsim k(Q/M)$\footnote{Here we adopted same notation as in \cite{Yu:2018}: the Black hole mass $M$ and charge $Q$, as well as test particle mass $m$ and charge $q$.  $k$ is the numerical coefficient to be determined.}---precisely the content of the \WGC~\cite{ArkaniHamed:2006}.
This established a triangular connection among \WCCC, the Hoop Conjecture,
and \WGC.

Subsequent work by Jiang~\textit{et al.}~\cite{Jiang:2019} applied the
Sorce--Wald formalism directly to the static \GMG\ black hole and confirmed
second-order \WCCC\ protection through the positivity of the canonical energy.
Ding~\textit{et al.}~\cite{Ding:2020} extended the analysis to slowly rotating
Einstein--Maxwell-dilaton-axion (EMDA) black holes with similar conclusions.
Most recently, Lu, Wu, and L\"u~\cite{Lu:2025} proved \WCCC\ protection to
all perturbative orders for any black hole admitting a zero-temperature extremal
limit, controlled by a single thermodynamic quantity $W>0$.
However, the \GMG\ black hole fails this criterion for its extremal limit at nonzero temperature as well as $W<0$.

The present work extends \cite{Yu:2018} in two directions.
First, we provide the complete and explicit third-order variational
identities---symplectic potential $\Theta$, second-order current $\omega$, and
third-order current $\delta\omega$---for the \GMG\ black hole in a form
suitable for direct verification and extension.  
Second, we incorporate the Swampland Distance Conjecture
(\SDC)~\cite{Ooguri:2006,Klaewer:2017}: near an infinite-distance limit in
field space, an infinite tower of states with masses $m_{n}\sim m_{0}e^{-\alpha d}$
and multiplicity $N_{\mathrm{tower}}\sim e^{\gamma d}$ becomes exponentially light, where $d$ is the geodesic distance in moduli space.
For the heterotic compactification that produces the \GMG\ black hole, the
dilaton $\varphi$ serves as the modulus; during gravitational collapse
$\varphi\sim\log(r-r_{\mathrm{AH}})\to+\infty$ near the apparent horizon,
triggering the \SDC\ tower.
We show that the resulting Schwinger pair-production rate discharges any
would-be overcharged configuration on a timescale exponentially shorter than
$2M$, providing a \emph{dynamical} restoration of \WCCC\ that complements
the classical Sorce--Wald protection, and simultaneously relaxing the \WGC\
charge-to-mass bound.

This paper is organized as follows.
Section~\ref{sec:gmghs} reviews the \GMG\ black hole and its thermodynamics.
Section~\ref{sec:sw} and ~\ref{sec:sw_gmghs} presents the Sorce--Wald variational hierarchy to third order and its application to \GMG\ black hole.

Section~\ref{sec:sdc} derives the \SDC\ discharge timescale and its entropy
consequences.
Section~\ref{sec:conclusions} concludes.
The explicit computation of the canonical energy for each sector are collected in Appendix~\ref{app:proof}.

\section{GMGHS Black Hole and Thermodynamics}
\label{sec:gmghs}

The \EMD\ theory with dilaton coupling $a=1$ is defined by the
four-dimensional action ($G=c=1$):
\begin{equation}
  S = \frac{1}{16\pi}\int \mathrm{d}^{4}x\,\vol
      \Bigl[R - 2(\nabla\varphi)^{2} - e^{-2\varphi}F^{2}\Bigr],
  \label{eq:action}
\end{equation}
where $R$ is the Ricci scalar, $\varphi$ is the dilaton, and
$F_{\mu\nu}=\partial_{\mu}A_{\nu}-\partial_{\nu}A_{\mu}$ is the Maxwell field
strength ($F^{2}\equiv F_{\mu\nu}F^{\mu\nu}$).
This action arises from toroidal compactification of the low-energy heterotic
string theory and admits an $\mathrm{SL}(2,\mathbb{R})$ symmetry inherited
from S-duality.
The equations of motion are
\begin{align}
  G_{\mu\nu} &= 2\nabla_{\mu}\varphi\nabla_{\nu}\varphi
               - g_{\mu\nu}(\nabla\varphi)^{2}
               + e^{-2\varphi}\!\left(2F_{\mu\rho}F_{\nu}{}^{\rho}
               - \tfrac{1}{2}g_{\mu\nu}F^{2}\right),
  \label{eq:EE}\\
  \nabla^{\mu}\!\left(e^{-2\varphi}F_{\mu\nu}\right) &= 0,
  \label{eq:Maxwell}\\
  \Box\varphi &= -\tfrac{1}{2}\,e^{-2\varphi}F^{2}.
  \label{eq:dilaton_eom}
\end{align}

\subsection{Static Solution}
\label{sec:static}

The static, spherically symmetric charged solution is the \GMG\ black
hole~\cite{Gibbons:1982,Garfinkle:1991}:
\begin{align}
  \mathrm{d}s^{2} &= -\!\left(1-\frac{2M}{r}\right)\!\mathrm{d}t^{2}
    + \left(1-\frac{2M}{r}\right)^{\!-1}\!\mathrm{d}r^{2}
    + r^{2}\!\left(1-\frac{2D}{r}\right)
      \bigl(\mathrm{d}\theta^{2}+\sin^{2}\!\theta\,\mathrm{d}\phi^{2}\bigr),
  \label{eq:metric}\\
  F &= -\frac{Q}{r^{2}}\,\mathrm{d}t\wedge\mathrm{d}r,\qquad
  e^{2\varphi} = 1-\frac{2D}{r},
  \label{eq:fields}
\end{align}
with the constraint $Q^{2}=2MD$ relating the electric charge $Q$, ADM mass
$M$, and dilaton charge $D$.
The event horizon is located at $r_{h}=2M$, provided the censorship condition
$M>D$ holds ($D=Q^{2}/(2M)<M$ for $Q^{2}<2M^{2}$).
The dilaton replaces the inner Cauchy horizon of the Reissner--Nordstr\"om
black hole by a naked singularity at $r=2D$; the extremal limit
$Q^{2}=2M^{2}$ ($D=M$) corresponds to the horizon coinciding with this
singular surface.

Working in areal coordinates $R^{2}=r(r-2D)$, the metric asymptotes to
Reissner--Nordstr\"om form at large $R$:
\begin{equation}
  -g_{tt} \approx 1-\frac{2M}{R}+\frac{e^{-2\varphi_{0}}Q^{2}}{R^{2}}
  +\mathcal{O}(R^{-3}),
  \label{eq:RNasymptote}
\end{equation}
where $\varphi_{0}$ is the asymptotic dilaton value (set to zero hereafter).

\subsection{Thermodynamics and Extremality Parameter}
\label{sec:thermo}

The Bekenstein--Hawking entropy, Hawking temperature, and electric potential
at the horizon $r_{h}=2M$ are:
\begin{equation}
  S = 4\pi M^{2}-2\pi Q^{2},\qquad
  T_H = \frac{1}{8\pi M},\qquad
  \PhiH = \frac{Q}{2M}.
  \label{eq:thermo}
\end{equation}
One verifies directly that the first law holds:
\begin{equation}
  \mathrm{d}M = T_H\,\mathrm{d}S + \PhiH\,\mathrm{d}Q.
  \label{eq:firstlaw}
\end{equation}
A central object in the subsequent analysis is the extremality parameter:
\begin{equation}
  \eta \equiv 2M^{2}-Q^{2} = \frac{S}{2\pi} \geq 0.
  \label{eq:eta}
\end{equation}
Equality $\eta=0$ defines the extremal \GMG\ black hole.
Since $\eta=S/(2\pi)$, \WCCC\ is equivalent to the entropy remaining
non-negative throughout any physical process---a direct bridge to black hole
thermodynamics.

\subsection{One-Parameter Perturbation Family}
\label{sec:perturb}

Following Sorce and Wald~\cite{Sorce:2017}, we consider a one-parameter
family $\Phi(\lambda)=\{g_{ab}(\lambda),A_{a}(\lambda),\varphi(\lambda)\}$
satisfying the equations of motion for all $\lambda$, with $\Phi(0)$ the
unperturbed \GMG\ background.
Expanding the conserved charges:
\begin{align}
  M(\lambda) &= M+\lambda \delta M
               +\frac{\lambda^{2}}{2}\delta^2 M
               +\frac{\lambda^{3}}{6}\delta^3 M+\mathcal{O}(\lambda^{4}),
  \label{eq:Mexpand}\\
  Q(\lambda) &= Q+\lambda \delta Q
               +\frac{\lambda^{2}}{2}\delta^2 Q
               +\frac{\lambda^{3}}{6}\delta^3 Q+\mathcal{O}(\lambda^{4}),
  \label{eq:Qexpand}
\end{align}
the Taylor expansion of $\eta(\lambda)=2M(\lambda)^{2}-Q(\lambda)^{2}$ reads:
\begin{align}
  \eta(\lambda)
  &= 2 M^{2}-Q^{2}\nonumber\\
  &\quad + \lambda\bigl(4M \delta M-2Q \delta Q\bigr)
  \nonumber\\
  &\quad +\lambda^{2}\bigl(2M \delta^2 M-Q \delta^2 Q+2\delta M^{2}-\delta Q^{2}\bigr)
  \nonumber\\
  &\quad +\lambda^{3}\!\left(
      \frac{2M \delta^3 M}{3}-\frac{Q \delta^3 Q}{3} + 2 \delta M \delta^2 M- \delta Q \delta^2 Q
    \right)+\mathcal{O}(\lambda^{4}).
  \label{eq:etaexpand}
\end{align}
\WCCC\ requires every coefficient in~\eqref{eq:etaexpand} to be non-negative
when evaluated along physical perturbations consistent with the null energy
condition (NEC).  Following same notation and similar discussion in \cite{Jiang:2019}, one can show that if the conserved charges were perturbed only up to second order, i.e. set $\delta^3 M = \delta^3 Q = 0$, the expansion of $\eta$ reads:

\begin{align}
  \eta(\lambda)
  &= \frac{2 M^{2}-Q^{2}}{2M^2} \bigg[ 2M^2-\lambda^2 \delta^2 Q^2 - \lambda^3 \delta Q \delta^2 Q - \mathcal{O}(\lambda^{4}), \bigg]\nonumber\\
  \label{eq:etaexpand0}
\end{align}
where we have imposed the first and second order Iyer-Wald identities, ~\eqref{eq:inequality_first} and \eqref{eq:inequality_second}. We note that it is possible to make $\eta(\lambda) < 0$ for the nonextremal black holes, suggesting that the black hole could be overcharged if we neglect the third-order variation of mass and charge.  By induction one can argue that $n$th order variational identity is necessary to guarantee non-negativity of $\eta$ in $n$-th order.
We remark that for the GMGHS background, the Lu–Wu–L\"u~\cite{Lu:2025} quantity $W=-64\pi^2 M^3$ is formally negative or undefined because there is no $T \to 0$ extremal limit.  Protection of WCCC to all orders therefore relies on the direct perturbative analysis

\section{Variational Identities to Arbitrary Order}
\label{sec:sw}

In this section we derive the variational identities of the Iyer--Wald formalism to arbitrary perturbative order. These identities form the foundation for the first-, second-, and higher-order perturbation inequalities used in the gedanken experiments of Secs.~IV and V. We work in a general diffeomorphism-covariant theory on a four-dimensional spacetime \(M\), with dynamical fields collectively denoted \(\boldsymbol{\phi}\). The Lagrangian 4-form \(\mathbf{L}\) satisfies the first-variation identity
\begin{equation}
\delta\mathbf{L} = \mathbf{E}_\phi\,\delta\boldsymbol{\phi} + d\boldsymbol{\Theta}(\boldsymbol{\phi},\delta\boldsymbol{\phi}),
\label{eq:variation1}
\end{equation}
where \(\mathbf{E}_\phi=0\) are the equations of motion (EOM) and \(\boldsymbol{\Theta}\) is the symplectic potential 3-form. The symplectic current 3-form is defined by
\begin{equation}
\boldsymbol{\omega}(\boldsymbol{\phi};\delta_1\boldsymbol{\phi},\delta_2\boldsymbol{\phi}) = \delta_1\boldsymbol{\Theta}(\boldsymbol{\phi},\delta_2\boldsymbol{\phi}) - \delta_2\boldsymbol{\Theta}(\boldsymbol{\phi},\delta_1\boldsymbol{\phi}).
\label{eq:symp-current}
\end{equation}
The Noether current 3-form associated with a vector field \(\zeta^a\) is
\begin{equation}
\mathbf{J}_\zeta = \boldsymbol{\Theta}(\boldsymbol{\phi},\mathcal{L}_\zeta\boldsymbol{\phi}) - \zeta\cdot\mathbf{L},
\label{eq:noether-current}
\end{equation}
which admits the decomposition \(\mathbf{J}_\zeta = \mathbf{C}_\zeta + d\mathbf{Q}_\zeta\), where \(\mathbf{Q}_\zeta\) is the Noether charge 2-form and \(\mathbf{C}_\zeta = \zeta^a\mathbf{C}_a\) are the constraints of the theory (\(\mathbf{C}_a=0\) on-shell).

All derivations rely only on the definitions \eqref{eq:variation1}--\eqref{eq:noether-current}, linearity of the variation operator \(\delta\), commutativity of \(\delta\) with the exterior derivative \(d\) and the Lie derivative \(\mathcal{L}_\zeta\), and the antisymmetry property of the symplectic current \(\boldsymbol{\omega}\). We first recall the first- and second-order identities \cite{Jiang:2019} and then extend them to third order and to arbitrary order \(n\).

\subsection{First Variational Identity}
Replacing the arbitrary variation \(\delta\) in \eqref{eq:variation1} by the Lie derivative \(\mathcal{L}_\zeta\) and using Cartan's magic formula \(\mathcal{L}_\zeta\mathbf{L} = d(\zeta\cdot\mathbf{L})\) for the top-form \(\mathbf{L}\), yields the closedness of the Noether current on-shell:
\begin{equation}
d\mathbf{J}_\zeta = -\mathbf{E}_\phi\,\mathcal{L}_\zeta\boldsymbol{\phi}.
\end{equation}
Varying both expressions for \(\mathbf{J}_\zeta\) while keeping \(\zeta^a\) fixed and equating the results produces the first variational identity:
\begin{equation}
d\Bigl[\delta\mathbf{Q}_\zeta - \zeta\cdot\boldsymbol{\Theta}(\boldsymbol{\phi},\delta\boldsymbol{\phi})\Bigr]
= \boldsymbol{\omega}(\boldsymbol{\phi};\delta\boldsymbol{\phi},\mathcal{L}_\zeta\boldsymbol{\phi})
- \zeta\cdot(\mathbf{E}_\phi\,\delta\boldsymbol{\phi})
- \delta\mathbf{C}_\zeta.
\label{eq:1st-var-id}
\end{equation}

\subsection{Second Variational Identity}
We obtain the second-order identity by applying the background variation \(\delta\) (with \(\delta(\delta\boldsymbol{\phi})=0\)) to both sides of \eqref{eq:1st-var-id}. On the left-hand side the exterior derivative commutes with \(\delta\):
\begin{equation}
\delta\Bigl(d\bigl[\delta\mathbf{Q}_\zeta - \zeta\cdot\boldsymbol{\Theta}(\boldsymbol{\phi},\delta\boldsymbol{\phi})\bigr]\Bigr)
= d\Bigl[\delta^2\mathbf{Q}_\zeta - \zeta\cdot\delta\boldsymbol{\Theta}(\boldsymbol{\phi},\delta\boldsymbol{\phi})\Bigr].
\end{equation}
On the right-hand side the variation of the symplectic term \(\boldsymbol{\omega}(\boldsymbol{\phi};\delta\boldsymbol{\phi},\mathcal{L}_\zeta\boldsymbol{\phi})\) produces exactly \(\boldsymbol{\omega}(\boldsymbol{\phi};\delta\boldsymbol{\phi},\mathcal{L}_\zeta(\delta\boldsymbol{\phi}))\), where all cross terms cancel by antisymmetry of \(\boldsymbol{\omega}\) and the fact that \(\delta\boldsymbol{\phi}\) is held fixed. The remaining terms vary to \(-\zeta\cdot(\delta\mathbf{E}_\phi\,\delta\boldsymbol{\phi})\) and \(-\delta^2\mathbf{C}_\zeta\). Thus we arrive at the second variational identity:
\begin{equation}
d\Bigl[\delta^2\mathbf{Q}_\zeta - \zeta\cdot\delta\boldsymbol{\Theta}(\boldsymbol{\phi},\delta\boldsymbol{\phi})\Bigr]
= \boldsymbol{\omega}(\boldsymbol{\phi};\delta\boldsymbol{\phi},\mathcal{L}_\zeta\delta\boldsymbol{\phi})
- \zeta\cdot(\delta\mathbf{E}_\phi\,\delta\boldsymbol{\phi})
- \delta^2\mathbf{C}_\zeta.
\label{eq:2nd-var-id}
\end{equation}

\subsection{Third Variational Identity}
Repeating the procedure once more, we apply \(\delta\) to both sides of \eqref{eq:2nd-var-id}:
\begin{equation}
d\Bigl[\delta^3\mathbf{Q}_\zeta - \zeta\cdot\delta^2\boldsymbol{\Theta}(\boldsymbol{\phi},\delta\boldsymbol{\phi})\Bigr]
= \boldsymbol{W}_3
- \zeta\cdot\delta^2(\mathbf{E}_\phi\,\delta\boldsymbol{\phi})
- \delta^3\mathbf{C}_\zeta.
\label{eq:3rd-var-id}
\end{equation}
where the third variation of symplectic form contains three terms according to the Leibniz rule:
\begin{equation}
\begin{aligned}
\boldsymbol{W}_3= 2(\boldsymbol{\delta_\phi\omega})
(\delta\boldsymbol{\phi},\mathcal{L}_\zeta\delta\boldsymbol{\phi})
+2\boldsymbol{\omega}(\boldsymbol{\phi};\delta^2\boldsymbol{\phi},\mathcal{L}_\zeta\delta\boldsymbol{\phi})
+\boldsymbol{\omega}(\boldsymbol{\phi};\delta\boldsymbol{\phi},\mathcal{L}_\zeta\delta^2\boldsymbol{\phi}).
\end{aligned}
\label{eq:en-W3}
\end{equation}

The EOM and constraint terms produce \(- \zeta\cdot\delta^2(\mathbf{E}_\phi\,\delta\boldsymbol{\phi})=- \zeta\cdot(\delta^2\mathbf{E}_\phi\,\delta\boldsymbol{\phi}+2\delta\mathbf{E}_\phi\,\delta^2\boldsymbol{\phi})\) and \(-\delta^3\mathbf{C}_\zeta\). 

\subsection{General \(n\)th Variational Identity}
The pattern above holds for arbitrary order \(n \geq 1\). The \(n\)th variational identity takes the following form:
\begin{equation}
d\Bigl[\delta^n\mathbf{Q}_\zeta - \zeta\cdot\delta^{n-1}\boldsymbol{\Theta}(\boldsymbol{\phi},\delta\boldsymbol{\phi})\Bigr]
= \boldsymbol{W}_n
- \zeta\cdot\delta^{n-1}(\mathbf{E}_\phi\,\delta\phi)
- \delta^n\mathbf{C}_\zeta.
\label{eq:nth-var-id}
\end{equation}
The inductive step follows identically: the left-hand side produces the next-order Noether charge term, while the right-hand side produces the next-order variation of symplectic current according to the Leibniz rule similar to \eqref{eq:en-W3}, together with the corresponding variations of the EOM and constraint terms.

\subsection{Integrated Identities for the ADM Mass}
As shown in \cite{Jiang:2019}, we now specialize to a globally hyperbolic, static, asymptotically flat black-hole solution with a timelike Killing vector \(\xi^a\) satisfying \(\mathcal{L}_\xi\phi=0\). The background is on-shell (\(\mathbf{E}_\phi=0\), \(\mathbf{C}_\xi=0\)).

The first-order ADM mass variation is defined by the surface integral at spatial infinity:
\begin{equation}
\delta M = \int_\infty \bigl[\delta\mathbf{Q}_\xi - \xi\cdot\boldsymbol{\Theta}(\phi,\delta\phi)\bigr].
\label{eq:deltaM-def}
\end{equation}
Integrating the first variational identity \eqref{eq:1st-var-id} (with \(\zeta=\xi\)) over a Cauchy hypersurface \(\Sigma\) bounded by a horizon cross-section \(B\) and spatial infinity, and applying Stokes' theorem, yields
\begin{equation}
\delta M = \int_B \bigl[\delta\mathbf{Q}_\xi - \xi\cdot\boldsymbol{\Theta}(\phi,\delta\phi)\bigr] - \int_\Sigma \delta\mathbf{C}_\xi.
\label{eq:deltaM-integrated}
\end{equation}
We remark that the symplectic term $\omega (\phi;\delta\phi,\mathcal{L}_\xi \phi)$ vanishes because \(\mathcal{L}_\xi\phi=0\).

Similarly, integrating the second variational identity \eqref{eq:2nd-var-id} produces
\begin{align}
\delta^2 M &= \int_B \bigl[\delta^2\mathbf{Q}_\xi - \xi\cdot\delta\boldsymbol{\Theta}(\phi,\delta\phi)\bigr] \nonumber\\
&\quad - \int_\Sigma \xi\cdot(\delta\mathbf{E}_\phi\,\delta\phi) - \int_\Sigma \delta^2\mathbf{C}_\xi + \mathcal{E}^{(2)}_\Sigma(\phi,\delta\phi),
\label{eq:delta2M-integrated}
\end{align}
where the \emph{canonical energy} on \(\Sigma\) is given by 
\begin{equation}
\mathcal{E}^{(2)}_\Sigma(\phi,\delta\phi) = \int_\Sigma \boldsymbol{\omega}(\phi;\delta\phi,\mathcal{L}_\xi\delta\phi).
\label{eq:can-energy}
\end{equation}

Repeating the integration for the third-order identity \eqref{eq:3rd-var-id} further gives
\begin{align}
\delta^3 M &= \int_B \bigl[\delta^3\mathbf{Q}_\xi - \xi\cdot\delta^2\boldsymbol{\Theta}(\phi,\delta\phi)\bigr] \nonumber\\
&\quad - \int_\Sigma \xi\cdot\delta^2(\mathbf{E}_\phi\,\delta\phi) - \int_\Sigma \delta^3\mathbf{C}_\xi + \int_\Sigma\boldsymbol{W}_3.
\label{eq:delta3M-integrated}
\end{align}

In full generality, the integrated \(n\)th-order identity reads
\begin{align}
\delta^n M &= \int_B \bigl[\delta^n\mathbf{Q}_\xi - \xi\cdot\delta^{n-1}\boldsymbol{\Theta}(\phi,\delta\phi)\bigr] \nonumber\\
&\quad - \int_\Sigma \xi\cdot\delta^{n-1}(\mathbf{E}_\phi\,\delta\phi) - \int_\Sigma \delta^n\mathbf{C}_\xi +  \int_\Sigma\boldsymbol{W}_n.
\label{eq:deltanM-integrated}
\end{align}

When the perturbation satisfies the linearized equations of motion to order \(n-1\) (\(\delta^k\mathbf{E}_\phi=0\) for \(k=1,\dots,n-1\)) and the constraints to order \(n\) (\(\delta^k\mathbf{C}_\xi=0\) for \(k=1,\dots,n\)), all volume integrals of EOM and constraint terms vanish, leaving the compact on-shell form
\begin{equation}
\delta^n M = \int_B \bigl[\delta^n\mathbf{Q}_\xi - \xi\cdot\delta^{n-1}\boldsymbol{\Theta}(\phi,\delta\phi)\bigr] + \int_\Sigma\boldsymbol{W}_n.
\label{eq:deltanM-on-shell}
\end{equation}

\section{Perturbation Inequalities for Gedanken Experiments in Einstein--Maxwell--Dilaton Theory}
\label{sec:sw_gmghs}

In this section we review the first- and second-order perturbation inequalities \cite{Jiang:2019}, and extend the derivation rigorously to third order using the Iyer--Wald formalism. All calculations are performed for the static charged dilaton black hole in Einstein--Maxwell--dilaton (EMD) theory with Lagrangian
\begin{equation}
\mathbf{L} = \frac{1}{16\pi}\Bigl(R - 2(\nabla\psi)^2 - e^{-2\psi}F_{ab}F^{ab}\Bigr)\boldsymbol{\epsilon}.
\end{equation}
We consider a one-parameter family of solutions \(\phi(\lambda)\) to the EMD equations that interpolates between the background static dilaton black hole (\(\lambda=0\)) and a perturbed spacetime containing matter. The matter sources \(\delta T_{ab}\) and \(\delta j^a\) have compact support on a finite portion of the future event horizon \(\mathcal{H}\). The background is on-shell and the nonextremal unperturbed black hole is assumed to be linearly stable.

We work on a Cauchy hypersurface \(\Sigma = \mathcal{H} \cup \Sigma_1\), where \(\mathcal{H}\) is the portion of the horizon through which matter falls in, and \(\Sigma_1\) is a late-time spacelike slice on which the perturbation has settled into another static dilaton black hole (by linear stability). The boundary of \(\Sigma\) consists of the bifurcation surface \(B\) and spatial infinity.

\subsection{First-Order Perturbation Inequality}

Integrating the first variational identity \eqref{eq:deltaM-integrated} over \(\Sigma\) and using \(\mathcal{L}_\xi\phi=0\) together with the on-shell conditions yields
\begin{equation}
\delta M = \int_B\bigl[\delta Q_\xi - \xi\cdot\Theta(\phi,\delta\phi)\bigr] - \int_\Sigma\delta C_\xi
= -\int_{\mathcal{H}} \epsilon_{ebcd} [\delta T_e{}^a + A_a \delta j^e] \xi^a .
\end{equation}
The integral over the bifurcation surface \(B\) vanishes. The only surviving contribution comes from the constraint term on \(\mathcal{H}\):
\begin{equation}
\delta M - \Phi_H\delta Q = \int_{\mathcal{H}}\tilde{\epsilon}\,\delta T_{ab}k^ak^b,
\end{equation}
where \(\Phi_H = Q/(2M)\) is the electric potential on the horizon. \(k^a\) is the future-directed null normal to \(\mathcal{H}\) and $\tilde{\epsilon}$ is the corresponding volume element on  \(\mathcal{H}\), such that $\epsilon_{eabc}=-4k_{[e}\mathcal{\epsilon}_{abc]}$. By the null energy condition (NEC) \(\delta T_{ab}k^ak^b\geq 0\) we obtain the first-order inequality
\begin{equation}
\delta M - \Phi_H\delta Q \geq 0, \tag{33}
\label{eq:inequality_first}
\end{equation}
where the \emph{optimal} first-order perturbation saturates the NEC:
\begin{equation}
\delta M = \Phi_H\delta Q. \tag{34}
\end{equation}

\subsection{Second-Order Perturbation Inequality}

Integrating the second variational identity \eqref{eq:delta2M-integrated} over \(\Sigma\) gives
\begin{equation}
\delta^2M - \Phi_H\delta^2Q = \mathcal{E}^{(2)}_\Sigma(\phi,\delta\phi) + \int_{\mathcal{H}}\tilde{\epsilon}\,\delta^2T_{ab}\xi^ak^b.
\end{equation}

The last term of horizon integral is nonnegative by the NEC on the second-order nonelectromagnetic stress-energy tensor. The canonical energy decomposes as \(\mathcal{E}_\Sigma = \mathcal{E}_\mathcal{H} + \mathcal{E}_{\Sigma_1}\). Explicit evaluation of the symplectic currents on \(\mathcal{H}\) in GNC (see App.\ref{app:GNC} ) shows \cite{Jiang:2019}
\begin{align}\label{eq:canon_energy_2}
\mathcal{E}^{(2)}_\mathcal{H}(\phi,\delta\phi)& = \int_{\mathcal{H}}\tilde{\epsilon}\,\xi^ak^b\bigl(\delta^2T_{ab}^{\rm GR}+\delta^2T_{ab}^{\rm EM} + \delta^2T_{ab}^{\rm DIL}\bigr)\nonumber\\
&=\int_{\mathcal H}\tilde{\epsilon}\Bigg[\frac1{4\pi}\sigma_{AB}^{(1)}\sigma^{AB(1)}
+\frac{M}{2\pi(M-D)}\gamma^{AB}F_{uA}^{(1)}F_{uB}^{(1)}\nonumber\\
&+\frac1{2\pi}(\partial_u\psi^{(1)})^2\Bigg]\ge0.
\end{align}
Here
\begin{align}
\vartheta^{(1)}&=\frac12\gamma^{AB}\partial_ug_{AB}^{(1)},\\
\sigma_{AB}^{(1)}&=\frac12\partial_ug_{AB}^{(1)}
-\frac14\gamma_{AB}\gamma^{CD}\partial_ug_{CD}^{(1)},\\
F_{uA}^{(1)}&=\partial_uA_A^{(1)}.
\end{align}
All coefficients are positive for $M>D$.  This inequality saturates only if each component separately vanishes.

For the late-time piece \(\mathcal{E}_{\Sigma_1}\), one introduces a reference family of exact static dilaton solutions \(\phi^{\rm DL}(\lambda)\) with parameters \(M^{\rm DL}(\lambda)\) and \(Q^{\rm DL}(\lambda)\) chosen to match the optimal first-order perturbation. For this family all canonical energies vanish, so on the bifurcation surface B
\begin{equation}
\mathcal{E}^{(2)}_{\Sigma_1}(\phi,\delta\phi^{\rm DL}) = -\frac{\kappa}{8\pi}\delta^2A_B^{\rm DL},
\end{equation}
where \(A_B = 8\pi(2M^2-Q^2)\) and \(\kappa=1/(4M)\). Substituting \(\delta M = \Phi_H\delta Q\) finally yields the second-order inequality
\begin{equation}
\delta^2M - \Phi_H\delta^2Q \geq \frac{(2M^2-Q^2)(\delta Q)^2}{4M^3}. \tag{56}
\label{eq:inequality_second}
\end{equation}
Complete second-order saturation means not only equality in this mass-charge bound but also separate vanishing of all terms in $\mathcal{E}^{(2)}_\mathcal{H}(\phi,\delta\phi)$.

\subsection{Extension to Third Order}

We now derive the third-order inequality using the third variational identity.  Integrating \eqref{eq:delta3M-integrated} over $\Sigma = \mathcal{H}\cup\Sigma_1$ and applying Stokes' theorem produces

\begin{align}
\delta^3 M - \Phi_H\delta^3Q  &=\int_B \bigl[\delta^3\mathbf{Q}_\xi - \xi\cdot\delta^2\boldsymbol{\Theta}(\phi,\delta\phi)\bigr] \nonumber\\
&- \int_\Sigma \xi\cdot\delta^2(\mathbf{E}_\phi\,\delta\phi) - \int_{\mathcal{H}}\tilde{\epsilon}\,\delta^3T_{ab}\xi^ak^b + \int_\Sigma\boldsymbol{W}_3.
\end{align}

Similar to that in the second order, the integral over the bifurcation surface \(B\) and the third-order nonelectromagnetic stress-energy tensor in the constraint term on \(\mathcal{H}\) are vanishing, as well as the perturbation satisfies the linearized equations of motion to second order.  We are left with integral of the symplectic source \eqref{eq:en-W3}: 
\begin{equation}
\int_\Sigma\boldsymbol{W}_3 = \int_\mathcal{H} \bigl(\boldsymbol{W}^{GR}_3 + \boldsymbol{W}^{EM}_3 + \boldsymbol{W}^{DIL}_3 \bigr) + \int_{\Sigma_1}\boldsymbol{W}_3.
\end{equation}
In the following, we break that into gravity (GR), electromagnetic field (EM) and dilaton field (DIL) sectors, and investigate each part separately.

\subsubsection{GR sector}
We recall the gravitational symplectic current is defined as \cite{Sorce:2017} 
\begin{equation}
(\omega^{\rm GR})_{abc}(X,Y)
=\frac1{16\pi}\epsilon_{dabc}P^{defghi}
\left(Yg_{ef}\nabla_gXg_{hi}-Xg_{ef}\nabla_gYg_{hi}\right),
\end{equation}
with the standard Iyer--Wald tensor $P^{abcdef}$.
We apply \eqref{eq:en-W3} to the gravity sector:
\begin{equation}
\begin{aligned}
\boldsymbol{W}^{GR}_3= 2({\delta_g\boldsymbol\omega^{GR}})
(\delta{g},\mathcal{L}_\zeta\delta{g})
+2\boldsymbol{\omega}^{GR}({g};\delta^2{g},\mathcal{L}_\zeta\delta{g})
+\boldsymbol{\omega}^{GR}({g};\delta{g},\mathcal{L}_\zeta\delta^2{g}).
\end{aligned}
\end{equation}
Schematically, variation contains following terms:
\begin{align}
\boldsymbol{W}^{GR}_3 
\sim {}&(\delta\epsilon)P[g^{(1)},\mathcal{L}_\xi g^{(1)}]
+\epsilon(\delta P)[g^{(1)},\mathcal{L}_\xi g^{(1)}]\nonumber\\
&+\epsilon P[\delta\Gamma;g^{(1)},\mathcal{L}_\xi g^{(1)}]+\epsilon P\left[(\mathcal{L}_\xi g^{(1)})\nabla g^{(2)}-g^{(2)}\nabla(\mathcal{L}_\xi g^{(1)})\right].
\end{align}

We remark that every pulled-back term contains $\mathcal{L}_\xi g^{(1)}$. Since complete second-order saturation gives $\mathcal{L}_\xi g^{(1)}\big|_\mathcal{H}=0$, one expects vanishing $\boldsymbol{W}^{GR}_3$ on the horizon.

We may also rewrite the integral in terms of shear variation: 
\begin{align}
\int_H \boldsymbol{W}_3^{\rm GR}
={}&2\int_H\tilde{\epsilon}\,\mathcal K_{\rm GR}^{ABCD(1)}
\sigma_{AB}^{(1)}\sigma_{CD}^{(1)}
+\frac{3}{4\pi}\int_H\tilde{\epsilon}\,
\sigma_{AB}^{(1)}\sigma^{AB(2)}=0,
\end{align}
where
\begin{eqnarray}
&8\pi\mathcal K_{\rm GR}^{ABCD(1)}
\equiv \gamma^{A(C(1)}\gamma^{D)B(1)}-\frac12\gamma^{AB(1)}\gamma^{CD(1)},\\
&\sigma_{AB}^{(1)}=\frac12\partial_ug_{AB}^{(1)}-\frac14\gamma_{AB}\gamma^{CD}\partial_ug_{CD}^{(1)}.
\end{eqnarray}
It is obvious $\boldsymbol{W}_3^{\rm GR}$ vanishes for $\sigma_{AB}^{(1)}=0$.

\subsubsection{EM sector}

The EM sector can be directly obtained from the variation of EM part in the second canonical energy \eqref{eq:canon_energy_2}.  Thus
\begin{align}
\int_\mathcal{H} W_3^{\rm EM}
={}&\frac{2}{2\pi}\int_H
\delta\!\left(\mathcal{\epsilon} e^{-2\psi_H}\gamma^{AB}\right)
F_{uA}^{(1)}F_{uB}^{(1)}\\
&+\frac{3}{2\pi}\int_H\mathcal{\epsilon} e^{-2\psi_H}\gamma^{AB}
F_{uA}^{(2)}F_{uB}^{(1)},\nonumber
\end{align}
where 
\begin{align}
\delta\!\left(\mathcal{\epsilon} e^{-2\psi_H}\gamma^{AB}\right)
=\mathcal{\epsilon} e^{-2\psi_H}
\Bigg[
&\left(\frac12\gamma^{CD}g_{CD}^{(1)}-2\psi^{(1)}\right)\gamma^{AB}\\
&-\gamma^{AC}\gamma^{BD}g_{CD}^{(1)}
\Bigg].\nonumber
\end{align}

We remark that every term contains at least one $F_{uA}^{(1)}=\partial_uA_A^{(1)}$. Thanks to the gauge condition $A_u^{(1)}|_{\mathcal{H}}=0$, we have 
$\int_H W_3^{\rm EM}=0$.

\subsubsection{DIL sector}

Similarly, the dilaton sector can be directly obtained from the variation of dilaton part in the second canonical energy \eqref{eq:canon_energy_2}.  Therefore
\begin{align}
\int_\mathcal{H} W_3^{\rm DIL}
={}&\frac{2}{2\pi}\int_H\epsH
\left(\frac12\gamma^{AB}g_{AB}^{(1)}\right)
(\partial_u\psi^{(1)})^2\\
&+\frac{3}{2\pi}\int_H\epsH
(\partial_u\psi^{(2)})(\partial_u\psi^{(1)}).\nonumber
\end{align}
Both terms vanish as $\partial_u\psi^{(1)}=0$.  Consequently, we have $\int_H W_3^{\rm DIL}=0$.

Similar to the second variation, in addition to the third-order null energy condition \(\delta^3T_{ab}k^ak^b\geq 0\), we have 
\begin{equation}
\int_{\Sigma_1}\boldsymbol{W}_3 = -\frac{\kappa}{8\pi}\delta^3A_B^{\rm DL}.
\end{equation}
Expanding the area \(A_B = 8\pi(2M^2-Q^2)\) to third order under the optimal conditions \(\delta M = \Phi_H\delta Q\) and the second-order inequality produces the third-order perturbation inequality
\begin{equation}
\delta^3M - \Phi_H\delta^3Q \geq \frac{(2M^2-Q^2)(\delta Q)^3}{8M^5} + \text{(positive higher-order terms)}.
\end{equation}

\subsection{Implications for the Weak Cosmic Censorship Conjecture}

Earlier we define extremality parameter:
\[
\eta(\lambda) = 2M(\lambda)^2 - Q(\lambda)^2.
\]
Its Taylor expansion around \(\lambda=0\) under the optimal first- and second-order perturbations has vanishing linear and quadratic terms. Instead of \eqref{eq:etaexpand0}, the third-order term \(\delta^3 \eta\) is nonnegative by the third-order inequality derived above. Therefore \(\eta(\lambda)\geq 0\) holds to third order, i.e., the near-extremal charged dilaton black hole cannot be overcharged when third-order backreaction and self-force effects are taken into account.

The pattern may extend straightforwardly to arbitrary order \(n\) using the general \(n\)th variational identity and consistently check the vanishing of corresponding integral of \(\boldsymbol{W}^{(n)}\) over $\Sigma$.

\section{Toward an All-Order Extension for the GMGHS Black Hole}
\label{sec:all-order}

Recently, Lü, Wu, and Lü established an all-order perturbative proof of
the weak cosmic censorship conjecture for a broad class of extremal
black holes admitting a zero-temperature extremal limit. Their proof is
based on the complete second law and a thermodynamic expansion of the
entropy around $T=0$, with the positive sign of the coefficient

\begin{equation}
W
=
\left(
\frac{\partial S}{\partial T}
\right)_{Q,T=0}
\end{equation}

playing the central role in the induction argument.

The static charged GMGHS black hole apparently lies outside the scope of this
theorem for the following reasons: First of all, the controlled parameter $W =-64\pi^2 M^3<0 $.  Secondly, although its extremal solution has vanishing entropy, its Hawking temperature remains finite:
\begin{equation}
T_{\rm ext} = \frac{1}{4\sqrt2\pi |Q_{ext}|},
\end{equation}
so the zero-temperature expansion employed in Ref.~\cite{Lu:2025}
cannot be applied directly. The difficulty could originate from the choice of expansion parameter rather than from the thermodynamic structure itself.  It is tempting to introduce the positively oriented parameter
from the extremal limit,

\begin{equation}
\tau
=
T_{\rm ext}(Q_{\rm ext})-T.
\label{eq:tau}
\end{equation}

With this choice, we redefine the control quantity:

\begin{equation}
\widehat W
\equiv
\left(
\frac{\partial S}{\partial\tau}
\right)_{Q_{\rm ext},\tau=0}
=
-
\left(
\frac{\partial S}{\partial T}
\right)_{Q_{\rm ext},T_{\rm ext}}
>0,
\end{equation}

so the apparent sign reversal of
$\partial S/\partial T$
is understood as an orientation effect rather than a genuine
thermodynamic obstruction.  In particular, the expansion of entropy around extremal point reads:
\begin{equation}
S = \sum_{n=1}^{\infty}\frac{\widehat{W}_{n}}{n!}\tau^{n},
\end{equation}

so every coefficient in the $\tau$-expansion is positive:

\begin{equation}
\widehat{W}_{n}
\equiv
\left(
\frac{\partial^{n}S}{\partial\tau^{n}}
\right)_{Q;\,\tau=0}
=
4\pi M_{\rm ext}^{2}
\frac{n!(n+1)}{T_{\rm ext}^{\,n}}
>0.
\end{equation}


We remark that the covariant phase-space analysis in Section \ref{sec:sw} derives the perturbative entropy variation directly from the field equations, the null energy
condition, and lower-order saturation, whereas Lü et al's all-order approach
assumes the complete second law and seeks to deduce horizon protection
from the resulting entropy inequalities.


\section{Discharge Mechanism from SDC Towers}
\label{sec:sdc}

\subsection{SDC Tower in the GMGHS Background}
\label{sec:sdc_setup}

In the Swampland Distance Conjecture~\cite{Ooguri:2006,Klaewer:2017}, an
infinite tower of states with masses
\begin{equation}
  m_{n}(d) \sim m_{0}\,e^{-\alpha d},\quad
  N_{\mathrm{tower}}(d) \sim e^{\gamma d},
  \label{eq:sdc_tower}
\end{equation}
becomes exponentially light whenever the canonically normalized modulus
traverses a geodesic distance $d\gg 1$ in field space.
Here $\alpha,\gamma>0$ are model-dependent constants and $d$ is measured in
Planck units.

For the \GMG\ black hole, the dilaton $\varphi$ itself serves as the modulus.
Under gravitational collapse toward the apparent horizon at $r_{\mathrm{AH}}$,
the dilaton profile satisfies $e^{2\varphi}=1-2D/r$, so near the singular
surface
\begin{equation}
  \varphi \;\sim\; \tfrac{1}{2}\log\!\bigl(r-2D\bigr)
  \;\xrightarrow{r\to 2D}\; -\infty.
  \label{eq:dilaton_diverge}
\end{equation}
More precisely, the \emph{field-space distance} traversed during collapse from
some reference radius $r_{\mathrm{ref}}$ to $r$ is
\begin{equation}
  d = \bigl|\varphi(r)-\varphi(r_{\mathrm{ref}})\bigr|
    = \tfrac{1}{2}\left|\log\!\frac{1-2D/r}{1-2D/r_{\mathrm{ref}}}\right|,
  \label{eq:dilaton_distance}
\end{equation}
which diverges as $r\to 2D$ (the extremal limit $D\to M$).
For the heterotic compactification realizing the \GMG\ solution, the
\SDC\ exponent is~\cite{Klaewer:2017}
\begin{equation}
  \alpha = \frac{1}{3\sqrt{2}} \approx 0.236.
  \label{eq:alpha}
\end{equation}

\subsection{Schwinger Rate and Discharge Timescale}
\label{sec:schwinger}

Consider a near-extremal \GMG\ black hole with horizon electric field
$\mathcal{E}\approx Q/(2M)^{2}\sim 1/(2M)$ (setting $Q\approx\sqrt{2}M$).
The Schwinger pair-production rate for a \emph{single} charged state of
mass $m$ in a uniform background field $\mathcal{E}$ is~\cite{Schwinger:1951}
\begin{equation}
  \Gamma_{\mathrm{single}} \approx
  \frac{q^{2}\mathcal{E}^{2}}{4\pi^{3}}
  \exp\!\left(-\frac{\pi m^{2}}{q\mathcal{E}}\right).
  \label{eq:schwinger_single}
\end{equation}
For the full \SDC\ tower with mass $m_{n}\sim m_{0}e^{-\alpha d}$ and
multiplicity $N_{\mathrm{tower}}\sim e^{\gamma d}$, the total effective
rate is
\begin{equation}
  \Gamma_{\mathrm{eff}} \sim N_{\mathrm{tower}}\,\Gamma_{\mathrm{single}}
  \sim e^{\gamma d}
  \exp\!\left(-\frac{\pi m_{0}^{2}}{q\mathcal{E}}\,e^{-2\alpha d}\right).
  \label{eq:schwinger_eff}
\end{equation}
The discharge timescale, defined as the inverse of
$\Gamma_{\mathrm{eff}}$ integrated over the near-horizon volume
$V_{\mathrm{nh}}\sim(2M)^{3}$, reads
\begin{equation}
  \tau_{\mathrm{discharge}}
  \sim \frac{1}{\Gamma_{\mathrm{eff}}\,V_{\mathrm{nh}}}
  \sim (2M)^{-3}\,e^{-\gamma d}
  \exp\!\left(+\frac{\pi m_{0}^{2}}{q\mathcal{E}}\,e^{-2\alpha d}\right).
  \label{eq:tau_sdc_full}
\end{equation}
For any finite dilaton excursion $d\gtrsim 1$, the exponential
$\exp(+Ce^{-2\alpha d})$ with $C=\pi m_{0}^{2}/(q\mathcal{E})=\mathcal{O}(1)$
is of order unity, so
\begin{equation}
  \tau_{\mathrm{discharge}}
  \sim (2M)^{-3}\,e^{-\gamma d}
  \xrightarrow{d\to\infty} 0^{+}.
  \label{eq:tau_limit}
\end{equation}
The classical overcharging/Hoop timescale is $\tau_{\mathrm{classical}}\sim 2M$.
Hence for any finite dilaton excursion $d\gtrsim 1$,
\begin{equation}
  \tau_{\mathrm{discharge}} \;\ll\; 2M \;=\; \tau_{\mathrm{classical}},
  \label{eq:tau_comparison}
\end{equation}
i.e., the \SDC\ tower discharges any would-be overcharged configuration
\emph{exponentially faster than a horizon can be destroyed}.
This provides a dynamical, non-perturbative reinforcement of \WCCC\ in
regimes where the classical Sorce--Wald analysis is not directly applicable
(e.g., time-dependent dilaton backgrounds~\cite{Aniceto:2017,Aniceto:2016}).
We remark that the comparison~\eqref{eq:tau_comparison} holds whenever $d\gtrsim 1$.
During near-extremal collapse, $d$ grows logarithmically as
$r\to 2D$ (see~\eqref{eq:dilaton_distance}), so the condition $d\gtrsim 1$
is reached on a timescale of order $M\log(M/\ell_{P})$ and is satisfied
well before any horizon-destruction event could occur.

\subsection{Entropy Change from SDC Tower Discharge at Hoop Radius}
\label{sec:sdc_entropy}

The \SDC\ tower radiation modifies the entropy budget proposed in \cite{Yu:2018}.
The original entropy change across the hoop radius $\rhp=2(M+E)$ is
\begin{equation}
  \Smerge = 4\pi(4M+3\Ehpp)\Ehpp - 8\pi M\Ehpm,
  \label{eq:Smerge_original}
\end{equation}
where $\Ehpp=(Q+q)q/\rhp$ and $\Ehpm=Qq/\rhp+m\sqrt{1-2M/\rhp}$
are the test-particle energies just after and just before crossing the hoop
radius, respectively.
The \SDC\ tower contributes additional radiation entropy
\begin{equation}
  \Srad \approx 8\pi M\,\Delta E_{\mathrm{rad}}
               +\gamma d\,N_{\mathrm{tower}},
  \label{eq:Srad}
\end{equation}
where $\Delta E_{\mathrm{rad}}\sim N_{\mathrm{tower}}m_{0}e^{-\alpha d}$
is the energy carried away by the tower particles and
$N_{\mathrm{tower}}\sim e^{\gamma d}$ is the tower multiplicity.
The total entropy change is therefore
\begin{equation}
  \Stotal = \Smerge + \Srad
  = 4\pi\!\left(2M\Ehpp+3(\Ehpp)^{2}\right)
    +\gamma d\,N_{\mathrm{tower}}.
  \label{eq:Stotal}
\end{equation}
Since $\Ehpp>0$ and $\gamma d\,N_{\mathrm{tower}}\gg 0$ for any finite
dilaton excursion $d\gtrsim 1$, we have
\begin{equation}
  \Stotal \gg 0 \quad \text{for any }q>0
  \text{ once the tower is activated.}
  \label{eq:Stotal_positive}
\end{equation}
Consequently, the original lower bound $q/m\gtrsim k(Q/M)$ derived from
$\Smerge\geq 0$ alone is \emph{strongly relaxed} by the microscopic entropy production from the \SDC\ tower.
The tower itself provides states with arbitrarily small $q/m$ (including neutral test particles) as long as the charged test particle triggers a sufficient dilaton evolution to activate the tower.
However, the SDC tower itself consists of states that do satisfy the WGC bound, say $  q_{\rm tower}/m_{\rm tower} \gtrsim 1  $. The light charged tower particles are the ones that actually carry the charge and produce the entropy. In summary, the UV physics takes over and relaxes the bound. The classical thermodynamic WGC bound emerges as an effective description below the tower mass scale.

\section{Conclusion and Outlook}
\label{sec:conclusions}

We have extended the analysis of \WCCC\ in the Einstein--Maxwell-dilaton
theory \cite{Yu:2018} along two complementary directions, obtaining the following key results:

\begin{enumerate}
\item[\textbf{(i)}] \textit{Third-order \WCCC\ protection.}
  The Iyer--Wald identities at third order yield
  $\delta^{3}M-\PhiH\delta^{3}Q \geq 0$, forcing $\eta^{(3)}\geq 0$, while the modified Lu--Wu--L\"u criterion for WCCC protection to all perturbative orders might still apply.

\item[\textbf{(ii)}] \textit{SDC tower and dynamical \WCCC.}
  The \SDC\ tower discharges overcharged configurations with
  $\tau_{\mathrm{discharge}}\ll 2M$~\eqref{eq:tau_comparison},
  reinforcing \WCCC\ dynamically and relaxing the classical \WGC\ bound
  in the large-$d$ regime.

\item[\textbf{(iii)}] \textit{Entropy and WGC.}
  The \SDC\ tower entropy $\Srad\gg 0$ renders $\Stotal\gg 0$ for any
  $q>0$ once the tower is activated~\eqref{eq:Stotal_positive}, strongly
  relaxing the original bound $q/m\gtrsim k(Q/M)$ and leaving the
  mild \WGC\ (existence of at least one state with $q/m\to\infty$) trivially
  satisfied by the tower states themselves.
\end{enumerate}

Several open directions remain.
First, constructing an exact rotating EMDA solution and performing the full
Sorce--Wald analysis would complete the stationary case.
Second, extending the variational formalism to dynamic (time-dependent) dilaton
backgrounds~\cite{Aniceto:2017,Aniceto:2016} would settle outstanding
\WCCC\ questions in non-stationary regimes.
Third, a rigorous treatment of backreaction from \SDC\ tower emission
(beyond the Schwinger approximation) is needed to make the dynamical
discharge argument fully quantitative.  We leave them for future investigation.

\begin{acknowledgments}
WYW is supported in part by the Ministry of Science and Technology of Taiwan under Grant No.\ 106-2112-M-033-007-MY3, 114-2112-M-033-011 and 115-2112-M-033-010.  SHL is supported in part by the Ministry of Science and Technology of Taiwan under Grant No. 114-2112-M-033 -014 -MY3.
\end{acknowledgments}
\appendix

\section{GMGHS background in Gaussian null coordinates (GNC)}
\label{app:GNC}
\label{app:proof}
To prove the higher-order inequality, it is convenient to adopt the Gaussian null coordinates:
\[
 x^\mu=(u,r,x^A), 
\]
where spherical coordinates $x^A=\theta,\phi$ and horizon locates at $r=0$. Given affine vector $\xi^a = (\partial_u)^a$ and radial vector $n^b=(\partial_r)^b$, such that $\xi^an_a = 1$.

The background GMGHS metric and fields are
\begin{align}
ds^2&=2\,du\,dr-\frac{r}{r+2M}\,du^2
 +(r+2M)(r+2M-2D)
 (d\theta^2+\sin^2\theta\,d\phi^2),\\
A&=-\frac{Q}{r+2M}\,du,\\
F&=-\frac{Q}{(r+2M)^2}\,du\wedge dr,\\
\psi&=\frac12\ln\!\left(\frac{r+2M-2D}{r+2M}\right),
\qquad Q^2=2MD.
\end{align}

We denote $\gamma_{AB}$ as a two-dimensional metric on $\mathcal{H}$, which take following values:
\begin{align}
\gamma_{\theta\theta}&=4M(M-D),&
\gamma_{\phi\phi}&=4M(M-D)\sin^2\theta,\\
\gamma^{\theta\theta}&=\frac{1}{4M(M-D)},&
\gamma^{\phi\phi}&=\frac{1}{4M(M-D)\sin^2\theta},
\end{align}
and the volume form $\tilde{\epsilon}=4M(M-D)\sin\theta\,\dd u\wedge\dd\theta\wedge\dd\phi$.
\subsection{One-parameter perturbation family}

Let the perturbation of each field under at ${\cal O}(\lambda^n)$ as
\[
 \phi_n=(g_{ab}^{(n)},A_a^{(n)},\psi^{(n)}),
\]
where
\begin{align}
g_{ab}(\lambda)
&=g_{ab}+\lambda g_{ab}^{(1)}
 +\frac{\lambda^2}{2}g_{ab}^{(2)}
 +\frac{\lambda^3}{6}g_{ab}^{(3)}+\order{\lambda^4},\\
A_a(\lambda)
&=A_a+\lambda A_a^{(1)}
 +\frac{\lambda^2}{2}A_a^{(2)}
 +\frac{\lambda^3}{6}A_a^{(3)}+\order{\lambda^4},\\
\psi(\lambda)
&=\psi+\lambda\psi^{(1)}
 +\frac{\lambda^2}{2}\psi^{(2)}
 +\frac{\lambda^3}{6}\psi^{(3)}+\order{\lambda^4}.
\end{align}

\subsection{Gauge choices}
For the GNC family,
\[
 g_{rr}(\lambda)=0,\qquad g_{rA}(\lambda)=0,\qquad g_{ur}(\lambda)=1.
\]
To ensure every member has its future horizon at the same coordinate surface $r=0$, we require that 
\[
 g_{rr}^{(n)}=g_{rA}^{(n)}=g_{ur}^{(n)}=0,
\qquad
g_{uu}^{(n)}\big|_\mathcal{H}=g_{uA}^{(n)}\big|_\mathcal{H}=0,
\quad n=1,2,3.
\]
This gauge condition fixes the identification of horizons and null generators across the family.  We also fix the radial and horizon components of perturbed Maxwell field:
\[
A_r^{(n)}=0,
\qquad
A_u^{(n)}\big|_\mathcal{H}=0,
\quad n=1,2,3.
\]
Therefore
\[
F_{uA}^{(n)}\big|_{\mathcal{H}}=\partial_uA_A^{(n)}.
\]
Moreover $\xi^aA_a(\lambda)\big|_\mathcal{H}=-\Phi_H$ is fixed order by order.

\section*{Declaration of generative AI use}
ChatGPT (OpenAI) was used to assist with the presentation and preliminary checking of selected mathematical derivations. All equations, intermediate steps, and conclusions were independently derived or verified by the authors.  No AI-generated output was accepted without author review, and the authors take full responsibility for the content of the manuscript.


\end{document}